\documentclass[]{spie}  

\usepackage{amsmath,amsfonts,amssymb}
\usepackage{graphicx}
\usepackage[colorlinks=true, allcolors=blue]{hyperref}
\usepackage{todonotes}
\usepackage{float}

\title{Heimdallr and Solarstein: alignment, calibration, and correction in the Asgard suite at the VLTI}
\author[a, *]{Adam K. Taras}
\author[a]{J. Gordon Robertson}
\author[b]{Josh Carter}
\author[a]{Fred Crous}
\author[b,c]{Benjamin Courtney-Barrer}
\author[b]{Grace McGinness}
\author[b]{Michael Ireland}
\author[a]{Peter Tuthill}
\affil[a]{Astralis-Usyd, Sydney Institute for Astronomy, School of Physics, University of Sydney, NSW 2006, Australia}
\affil[b]{Research School of Astronomy and Astrophysics, Australian National University, Canberra 2611, Australia}
\affil[c]{European Southern Observatory, Alonso de Córdova 3107 Vitacura, 19001, Santiago, Chile}

\authorinfo{adam.taras@sydney.edu.au}

\usepackage{acronym}		
\usepackage{glossaries}		

\usepackage{booktabs}

\newacronym{USyd}{USyd}{the University of Sydney}
\newacronym{ANU}{ANU}{Australia National University}
\newacronym{ESO}{ESO}{European Southern Observatory}

\newacronym{NSF}{NSF}{National Science Foundation}
\newacronym{LIEF}{LIEF}{Linkage Infrastructure, Equipment and Facilities}

\newacronym{VLTI}{VLTI}{Very Large Telescope Interferometer}
\newacronym{ATs}{ATs}{auxiliary telescopes}
\newacronym{UTs}{UTs}{unit telescopes}
\newacronym{ELT}{ELT}{Extremely Large Telescope}

\newacronym{STS}{STS}{six telescope simulator}
\newacronym{GPAO}{GPAO}{GRAVITY+ adaptive optics}

\newacronym{WFS}{WFS}{wavefront sensor}
\newacronym{ADC}{ADC}{atmospheric dispersion corrector}
\newacronym{LDC}{LDC}{longitudinal dispersion corrector}
\newacronym{AO}{AO}{adaptive optics}
\newacronym{SNR}{SNR}{signal to noise ratio}

\newacronym{LED}{LED}{light emitting diode}

\newacronym{OAP}{OAP}{off-axis paraboloid}
\newacronym{DM}{DM}{deformable mirror}
\newacronym{MEMS}{MEMS}{micro-electromechanical system}
\newacronym{IR}{IR}{infrared}

\newacronym{CAD}{CAD}{computer aided design}

\begin{document} 
\maketitle

\begin{abstract}
The Asgard instrument suite proposed for the ESO’s Very Large Telescope Interferometer (VLTI) brings with it a new generation of instruments for spectroscopy and nulling. Asgard will enable investigations such as measurement of direct stellar masses for Galactic archaeology and direct detection of giant exoplanets to probe formation models using the first nulling interferometer in the southern hemisphere. We present the design and implementation of the Astralis-built Heimdallr, the beam combiner for fringe tracking and stellar interferometry in K band, as well as Solarstein, a novel implementation of a 4-beam telescope simulator for alignment and calibration. In this update, we verify that the Heimdallr design is sufficient to perform diffraction-limited beam combination. Furthermore, we demonstrate that Solarstein presents an interface comparable to the VLTI with co-phased, equal intensity beams, enabling alignment and calibration for all Asgard instruments. In doing so, we share techniques for aligning and implementing large instruments in bulk optics.
\end{abstract}


\section{Introduction}
The \gls{VLTI} is world-leading observatory that regularly enables tremendous leaps in our understanding of astronomy and astrophysics \cite{kervella_vinci_2000, petrov_amber_2007,gravity_collaboration_detection_2018}. The Asgard instrument suite\cite{martinod_high-angular_2023} is a \gls{VLTI} visitor instrument in development, bringing a new generation of instruments that will leverage long baseline interferometry to perform spectroscopy and nulling. 

Heimdallr, named after the Norse god with keen senses, provides the wavefront corrections through deformable mirrors, provides fringe tracking and low-order wavefront sensing as well as visibility science. Solarstein, named after the stones believed to have been used by Vikings to navigate, is a four beam telescope simulator. It provides an interface to the rest of Asgard very similar to that of the \gls{VLTI}, enabling alignment and daytime calibration. The detailed designs and requirements of these modules are documented in previous work \cite{taras_heimdallr_2024}. 

These modules are under construction for lab testing, and this paper provides an update on the progress. Specifically, the contributions of this work are:
\begin{itemize}
    \item to quantify and test the lab implementation of Solarstein and parts of Heimdallr, sharing how we meet the design requirements \cite{taras_heimdallr_2024}; and
    \item to draw generalisable lessons from this process that are applicable to current instruments.
\end{itemize}

\section{Implementation progress}

\subsection{Solarstein}

We have implemented a near-complete version of this novel simulator, excluding only the spectral calibration source.

\begin{figure}[h]
    \centering
    \includegraphics[width=0.99\textwidth]{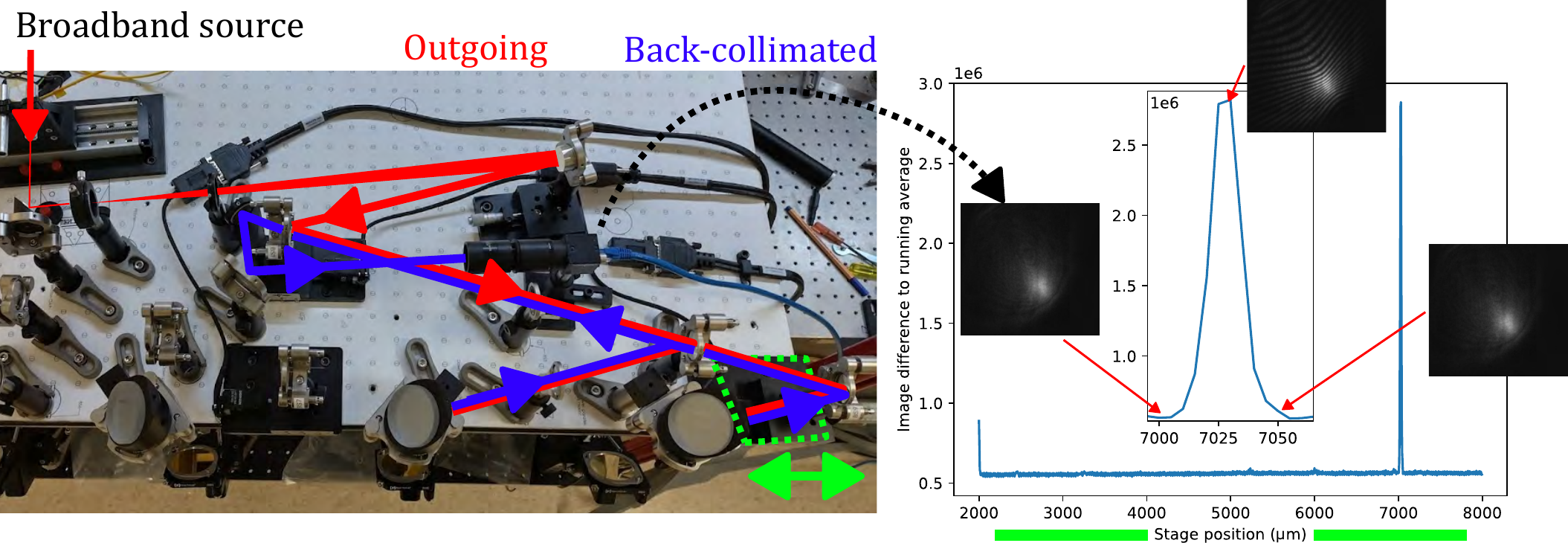}
    \caption{Co-phasing of Solarstein, using beams 3 and 4 as an example. In the lab setup (left) the outgoing (red) beams are back-collimated (blue) and additional fold mirrors are used to place the resulting beam on a detector. The motorised stage for the beamsplitter (green) is then moved and images recorded. The scan is analysed (right) by looking at the difference of the current image to a rolling average, and fringes are found on a broadband source within the expected tolerance of the setup.}
    \label{fig:solarstein-fringes}
\end{figure}

An annotated image of Solarstein on the upper level is shown in \autoref{fig:solarstein-fringes}, also summarising the results for a pairwise co-phasing between beams 3 and 4 (numbered left to right). After initial alignment with a laser, mirrors are placed on the lower level at the same distance along the beam to within screw-hole tolerances and the light reflected back through the beamsplitter array and onto a detector. The simulator is co-phased using a red superluminescent diode (650nm, Thorlabs part number: SLD650B) with a 6nm bandwidth. We take steps on the linear stage to change the delay of beam 4 relative to beam 3 and capture a series of images. To find the location of the fringes, we calculate the image difference $\Delta$ of the image at index $i$ to a rolling average of previous images, given by
    \begin{align}
        \Delta(i) = \sum_{\mathrm{pixels}} (I_i - \bar{I})^2,\ \ \        \bar{I} = \mathrm{avg}(\{I_{i-1}, I_{i-2} \dots I_{\max(i-N,0)}\})
    \end{align}
where $I_j$ is the image at index $j$ and $N=20$ is the number of images to perform the average over. This direct calculation is fast (not requiring a Fourier transform), agnostic to the unknown direction of the fringes and robust to slow changes in the image throughout a wide sweep. The resulting signal (\autoref{fig:solarstein-fringes}, right) is extremely clear and indicates the position the stage should be set to for broadband testing. We repeat this process for two other pairwise combinations to co-phase all four beams. 

The beamsplitters in the array are custom wedged CaF$_2$ windows coated with 6nm nickel. When viewing from the coated side, the optic functions as intended. From the reverse side, however, the behaviour is far from symmetric, instead producing two reflections of roughly equal intensity from the front and back surfaces at visible wavelengths. This poses a challenge for co-phasing beams 2 and 3, where the back-collimated beams are not symmetric, with beam 2 undergoing an extra reflection from the back surface of the first beamsplitter in the array versus a transmission only for beam 3. This produces a noticeable drop in visibility for the back-collimated beam but the peak in the fringe search still has a signal-to-noise ratio of order $10^2$. We also found that the soft nature of the CaF$_2$ substrate renders it incompatible with standard flexure based retention, instead requiring glue-in mounts.

We have also verified that the beams to the lower level meet the requirements on beam diameter, illumination uniformity and repeatability after remotely switching sources.

Finally, we have also verified the throughput of the thermal source (Thorlabs part number: SLS201/M) on a Pointgrey detector at the lower level. We imaged the source onto approximately 1000 pixels and saturated an IMX265 sensor and found it saturated at approximately 80\,ms. Accounting for the sensor properties yields a throughput of $5\times 10^8$ photons/s/beam at 800nm. This is sufficiently bright for Asgard instruments to use even at longer wavelengths and is brighter than astrophysical sources.

\subsection{Heimdallr}

We have implemented 3/4 of the beams of Heimdallr up to the intermediate focus plane, including one beam with a deformable mirror (DM). Rather than present detailed tests, this subsection shares our progress and lessons learned in the alignment process. 

\begin{figure}[h]
    \centering
    \includegraphics[width=0.65\textwidth]{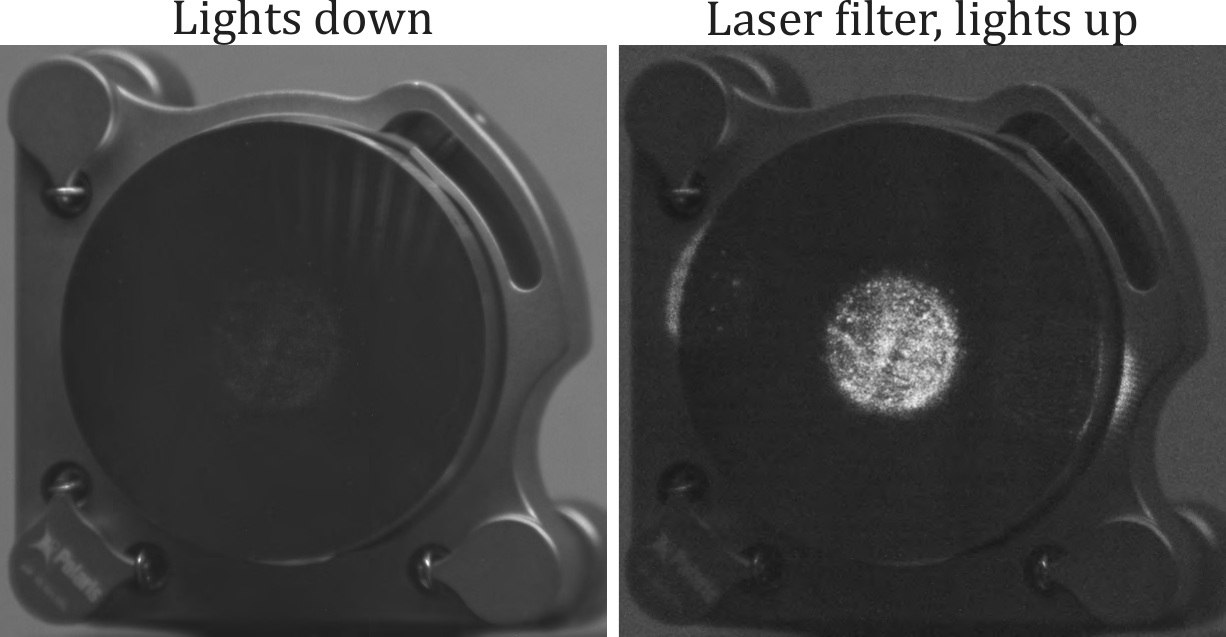}
    \caption{Alignment by observing scattering on the first surface of Heimdallr. At the lower level, the input beam is dim such that the lights must be extremely dimmed to view the beam. (left) Using a camera still requires lights to be dimmed to barely see the scatting profile. (right) We instead employ a laser line filter and are able to work with blue \glspl{LED} on full brightness. The resulting sensitivity enables alignment with no parallax or target offset error.}
    \label{fig:heimdallr_camera}
\end{figure}

Once at the lower level, the visible alignment laser can become faint. Instead of increasing laser power or working in extreme darkness -- which poses risks with expensive optics -- we employ a laser line filter on a camera, performing alignment by observing the scattering off a target surface. This is shown in \autoref{fig:heimdallr_camera}. Our approach also eliminates parallax errors from using targets, keeps all lasers at an eye safe level at the upper level and enables alignment with only a single person.

\begin{figure}[H]
    \centering
    \includegraphics[width=0.65\textwidth]{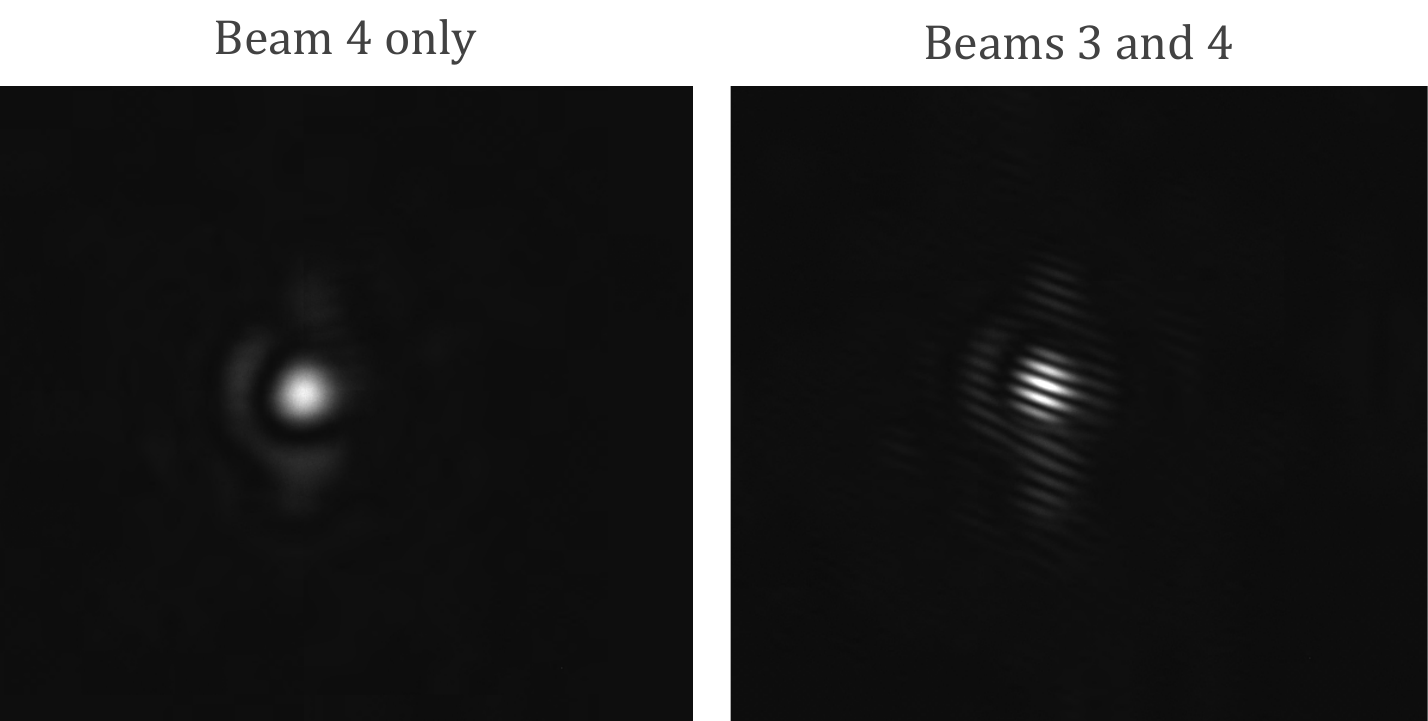}
    \caption{Beam quality at the intermediate detector at $\lambda=635$\,nm. (left) The image from a single beam is close to an Airy ring without any DM correction, validating that the design and alignment enables diffraction limited performance in K band. (right) Fringes between two beams, one of which has a deformable mirror.}
    \label{fig:beam_quality_and_fringes}
\end{figure}

The Heimdallr design includes an intermediate focus before the Narcissus box. \autoref{fig:beam_quality_and_fringes} illustrates some preliminary results in laser light at this focus from two of the beams. The images demonstrate that, in K band, we can expect diffraction limited performance. Also, despite scattering from the DM changing the intensity, the fringes are still high visibility and will be sufficient for testing the Heimdallr control loop.

\begin{figure}[h]
    \centering
    \includegraphics[width=0.95\textwidth]{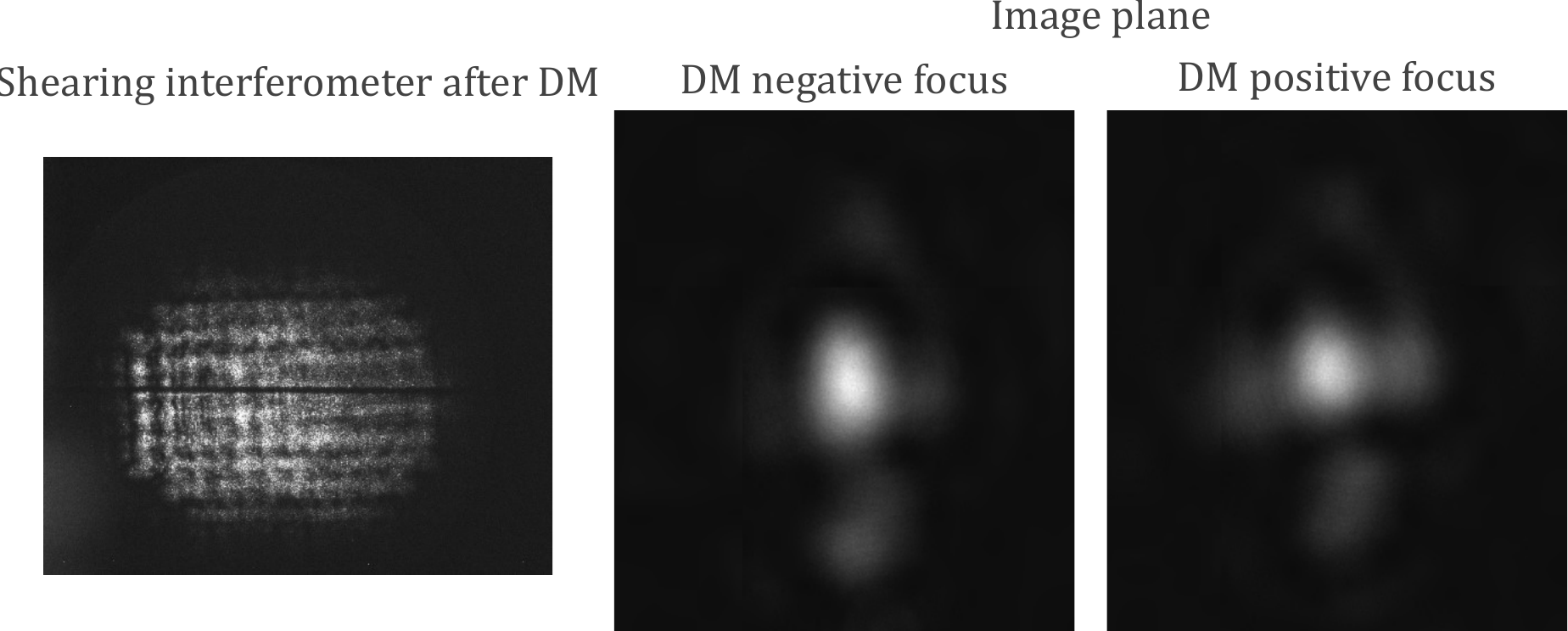}
    \caption{Challenges of collimating through a DM. (left) Camera image of the shearing interferometer after beam collimation through the DM. The scattering from the mirror renders the shearing interferometer incapable of sensing the collimation of the beam. (middle, right) Detector image at intermediate focus with correct collimation, with different focus aberrations applied to the DM and approximately equal Strehl ratio in either case.  }
    \label{fig:heimdallr_dm_shear}
\end{figure}

The addition of the DM in the setup complicates the alignment since the scattering from the surface confounds the signal from the shearing interferometer, as shown on the left in \autoref{fig:heimdallr_dm_shear}. We are instead able to infer the collimation by actively probing the beam with the DM. Assuming that the intermediate focus and the spherical focus mirror is correct (with large tolerances due to the 4m radius of curvature), applying a defocus alternating between positive and negative values changes the image asymmetrically when the beam is not collimated. The collimation is adjusted such that the Strehl ratio is unchanged between the positive and negative focus.

We can also present that the manufacturing of the critical Narcissus mirrors was successful. In particular, the tolerances on the positions of the holes was 50\,$\mu$m, and the errors in hole positions measured ranged from 6 to 22\,$\mu$m. This will enable a reduction of the thermal background on the detector, with each pixel either seeing the cold stop or the sky rather than the glowing (in K band) lab.

\section{Conclusion and future work}

We present an update on the implementation of Heimdallr: a beam combiner for fringe tracking and visibility science; and Solarstein: a novel four beam telescope simulator. These are critical modules, and part of the wider Asgard suite in development as a visitor instrument for the \gls{VLTI}. For Solarstein, our lab setup meets the requirements from the original design, providing four co-phased, equal intensity beams that will enable alignment and daytime calibration of the entire Asgard suite at Mt. Paranal. We also describe the status of Heimdallr and share our method of aligning the instrument with a dim alignment laser whilst maintaining safe working conditions.


\acknowledgments 
 
We acknowledge support from Astralis - Australia’s optical astronomy instrumentation Consortium - through the Australian Government’s National Collaborative Research Infrastructure Strategy (NCRIS) Program. This work used the SA and ANU nodes of the NCRIS-enabled Australian National Fabrication Facility (ANFF).

This research was supported by the Australian Research Council (ARC) Linkage Infrastructure Funding (LIEF) grant LE220100126. 

\bibliography{references} 
\bibliographystyle{spiebib} 

\end{document}